\documentclass[rpm,twocolumn,letterpaper]{revtex4}

\usepackage{graphicx}
\usepackage{amsfonts}
\usepackage{psfrag}

\newcommand{\RE}{\mbox{Re}}
\newcommand{\IM}{\mbox{Im}}

\newcommand{\be}{\begin{equation}}
\newcommand{\ee}{\end{equation}}
\newcommand{\beq}{\begin{eqnarray}}
\newcommand{\eeq}{\end{eqnarray}}
\newcommand{\pd}[2]{\frac{\partial #1}{\partial #2}}
\newcommand{\mat}[4]{\left(\begin{array}{cc}
 #1& #2\\
 #3&#4
 \end{array}\right)}

\begin{document}

\title{Dramatic Shape Sensitivity of Directional Emission Patterns from
Similarly Deformed Cylindrical Polymer Lasers}

\author{Harald~G.~L.~Schwefel, Nathan~B.~Rex, Hakan~E.~Tureci, Richard~K.~Chang and A.~Douglas~Stone}
\affiliation{Yale University, Department of Applied Physics, \\P.O.~Box 208284, New Haven, CT 06520-8284, USA}
\author{Joseph Zyss}
\affiliation{Ecole Normale Sup\'erieure de Cachan, \\94235 Cachan, Cedex, France}

\begin{abstract}
Recent experiments on similarly shaped polymer micro-cavity lasers show a dramatic difference in the far-field emission patterns. We show for different deformations of the ellipse, quadrupole and hexadecapole that the large differences in the far-field emission patterns is explained by the differing ray dynamics corresponding to each shape. Analyzing the differences in the appropriate phase space for ray motion, it is shown that the differing geometries of the unstable manifolds of periodic orbits are the decisive factors in determining the far-field pattern. Surprisingly, we find that strongly chaotic ray dynamics is compatible with highly directional emission in the far-field.
\end{abstract}
\maketitle

\section{Introduction}
\label{section:int}
Spherical, cylindrical and disk-shaped dielectric cavities have been of interest as compact, high-Q optical resonators to be used in micro-lasers and integrated optics applications\cite{chang_book,yamamoto93}. The high-Q modes of such devices are ``whispering gallery modes'' (WG) which circulate inside the boundary confined by total internal reflection. Due to their intrinsic rotational symmetry such devices require additional symmetry-breaking elements to couple in a directional manner and to optimize the Q-values for a given application. Some time ago it was shown by N\"ockel, Stone and Chang\cite{NockelSC94,NockelSCGC96,NockelS97} that smooth deformations of such spherically symmetric cavities (termed ARCs for {\em asymmetric resonant cavities}) still have anisotropic WG modes which now have intrinsically directional emission and Q-values which were tunable by the degree of deformation introduced. The optical physics of such resonators is non-trivial and interesting because the ray dynamics in such a case is partially chaotic and standard real-space ray-tracing is not helpful in analyzing their properties. Instead, phase space methods taken from non-linear classical dynamics such as the surface-of-section method (see below) allowed a much clearer picture of the physics, leading to qualitative predictions for the high intensity emission directions from quadrupole-deformed ARCs\cite{NockelSCGC96,NockelS97,ChangCSN00}. This phase space picture and associated ray simulations were shown to agree semi-quantitatively with exact numerical calculations of the linear resonances of quadrupole ARCs\cite{NockelS97,ChangCSN00}.

\begin{figure}[!t]
\centering
\includegraphics[width=\linewidth]{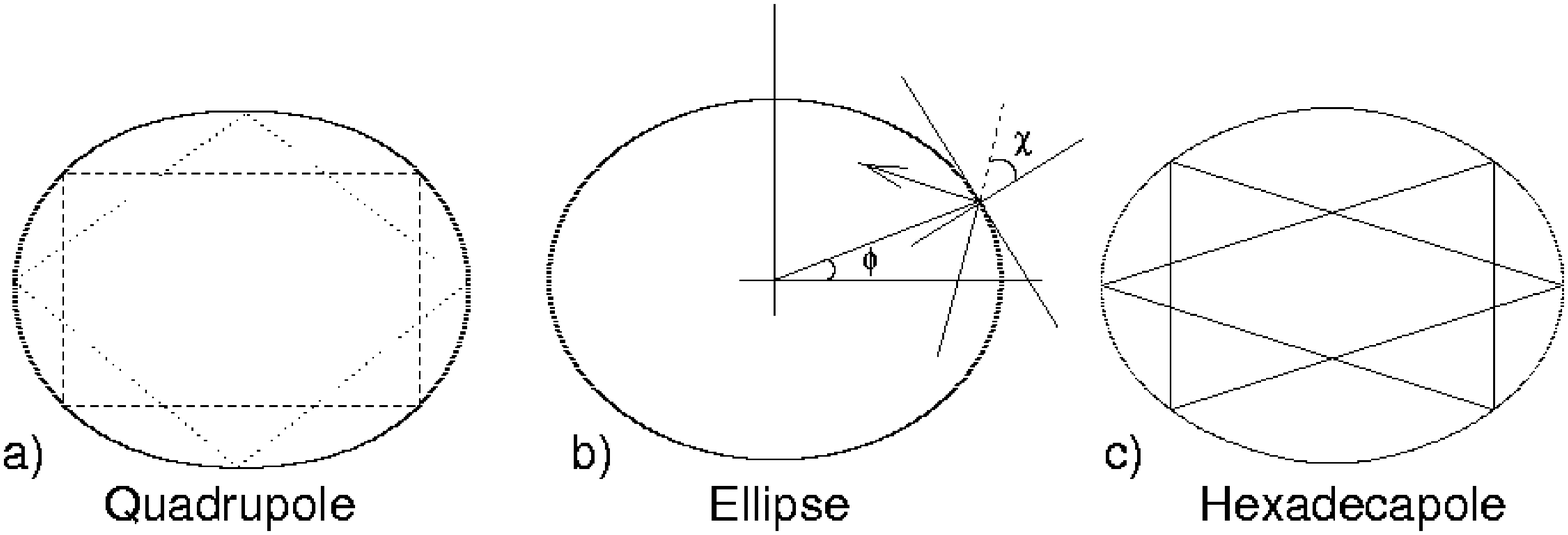}
\caption{Cross-sectional shapes of micro-pillar resonators studied: a) The quadrupole, defined in polar coordinates by $R=R_0(1+\varepsilon\cos 2\phi)$, b) The ellipse, defined by $R=R_0(1+((1+\varepsilon)^4-1)\sin^2\phi)^{-1/2}$ and c) The Quadrupole-Hexadecapole, defined by $R=R_0(1+\varepsilon(\cos^2\phi+\frac{3}{2}\cos^4\phi))$ all at a deformation of $\varepsilon=0.12$. Note that all shapes have horizontal and vertical reflection symmetry and have been defined so that the same value of $\varepsilon$ corresponds to approximately the same major to minor axis ratio. $\chi$ (Fig.~b) is the angle of incidence of a ray with respect to the local normal. In a) and c) we show short periodic orbits (``diamond, rectangle, triangle") relevant to the discussion below.}
\label{fig:shapes}
\end{figure}

The essence of the N\"ockel-Stone ray model for ARCs was that emission from deformed WG modes could be viewed as refractive escape of rays which were initially trapped by total internal reflection, but which, due to their chaotic dynamics, could diffuse chaotically in angle of incidence, $\chi$, until they reach the critical angle, $\sin \chi_c = 1/n$ ($n$ is the index of refraction of the resonator, assumed to be surrounded by air) and refracted out. One might naively assume that such chaotic ray dynamics would generate a fluctuating and pseudo-random emission pattern, but in fact the ray motion follows a dominant flow pattern in the phase space favoring escape at certain points on the boundary and in certain directions in the far-field. Both numerical experiments and later experimental measurements on ARC lasers found highly directional emission patterns in the far-field. For the quadrupole and related ARCs it was argued that the flow pattern was approximately describable as rapid motion along adiabatic invariant curves (which could be calculated from knowledge of the boundary shape) and slow diffusion in the transverse direction. However major deviations from the flow pattern would occur in the vicinity of stable periodic ray orbits for reasons to be discussed in detail below. As the shape of the adiabatic curves and the location of stable and unstable periodic orbits is quite sensitive to boundary shapes, this theory predicted a rather dramatic sensitivity of the emission patterns from ARC micro-lasers to the shape of the boundary. In the current work we explore a range of interesting ARC shapes, both within and outside the range of earlier theory. One of our goals is to test the earlier theoretical predictions experimentally for the first time with a controlled series of boundary shapes for polymer ARC microlasers. The shapes we consider are quadrupole, elliptical and hexadecapole ARCs (see definitions, caption of Fig.~\ref{fig:shapes}). We find that; 1)~There is a remarkable and reproducible difference in the lasing emission patterns from ARC lasers with very similar boundary shapes. 2)~The basic difference between chaotic (quadrupole) and non-chaotic (elliptical) ARC emission patterns is in agreement with the predictions of N\"ockel and Stone based on the adiabatic model. 3)~Nonetheless the {\em persistence} of highly directional emission patterns for highly deformed quadrupolar ARCs is inconsistent with the adiabatic model and is a quite surprising experimental discovery. 4)~A new theoretical model is proposed which attributes the high emission directions observed for the chaotic shapes to the flow pattern produced by the unstable manifolds of short periodic orbits; this flow pattern differs significantly from the adiabatic model. This model is shown to explain both the persistence of narrow emission peaks in the quadrupole at high deformation and the major shift in emission directionality at large deformation for ARCs with hexadecapole deformation. It also predicts that completely chaotic boundary shapes, such as the stadium, can nonetheless exhibit highly directional emission.

Further below (section~\ref{section:surpise}) we will review the basic ideas of mixed phase space, which describes systems with a mixture of chaotic and regular dynamics. At that point we will explain in detail the adiabatic model of N\"ockel and Stone and how it fails to account for the emission data for the highest deformations of resonators with mixed phase space. However we review here just a few basic concepts to put the current experimental and theoretical work in perspective. At maximal radial deformations of order 15\% from a reference circle the quadrupole and hexadecapole resonators have primarily chaotic ray dynamics and this implies that most initial conditions corresponding to total internal reflection (TIR) belong to ray trajectories which eventually will strike the boundary below TIR and hence will escape rapidly by refraction. The prior model posited a correspondence between a set of totally-internally-reflected initial conditions on rays and WG modes of the deformed resonator. This correspondence, based on adiabatic invariants, will be described below. The correspondence is essentially exact for the ellipse but is only a rough approximation for the quadrupole. Once ray initial conditions are chosen, the emission patterns can be calculated using straightforward ray simulations as described in\cite{NockelSCGC96,NockelS97} and a qualitative understanding of the emission patterns is possible based on properties of the phase-space flow. Using this picture of phase space flow, N\"ockel and Stone predicted\cite{NockelSCGC96,NockelS97, ChangCSN00}; that quadrupole resonators with deformations in the range of 10-12\% and index of refraction $n=1.5$ would emit very differently than elliptical resonators with the same major to minor axis ratio. (Similarly, but less relevant here, they showed that a $n=2$ quadrupole resonator of exactly the same shape would emit differently from the $n=1.5$ case). Specifically the elliptical resonators would emit from the points of highest curvature on the boundary roughly in the tangent direction ($90^{\circ}$) while the quadrupoles of the same index would emit at roughly a $35^{\circ}-45^{\circ}$ angle to the major and minor axes. It was argued that the origin of this effect is the presence in the quadrupole of a stable four-bounce periodic ray orbit which prevents emission from the highest curvature points in the tangent direction, an effect they termed ``dynamical eclipsing"\cite{NockelSCGC96,NockelS97}. This finding was supported by numerical solutions of the linear wave equation for the quasi-bound states and their far-field emission patterns. Mode selection and non-linear lasing processes were not treated in the theory. A subsequent experiment on lasing droplets by Chang et al.\cite{ChangCSN00} was successfully interpreted as strong evidence for such dynamically eclipsed lasing modes. However this experiment was less direct than desirable for two reasons. First, the droplet was a deformed sphere with many possible lasing modes, most of which were not of the two-dimensional type considered in references\cite{NockelSCGC96,NockelS97}; it was argued, but not experimentally shown, that the 2-d ``chaotic" modes dominated the lasing emission. Second, the droplet shape deformation was not controlled and could not be manipulated to turn the effect on and off. Recently, Lacey et al.\cite{nockel03} reported an experiment on nearly spherical resonators where they addressed the former but not latter problem. The experiments reported here remedy both of these shortcomings. First the lasers are deformed cylinders and the lasing modes are truly two-dimensional. Second we have fabricated the boundary shapes using a mask to conform to the desired cross-sectional profile. Hence we can directly compare e.g.\ quadrupolar and elliptical ARC lasers and observe the presence or absence of the dynamical eclipsing effect over a range of deformations. 

\section{Experimental Data}
\label{section:exp}
\begin{figure*}[!hbt]
\centering
\includegraphics[width=0.7\linewidth]{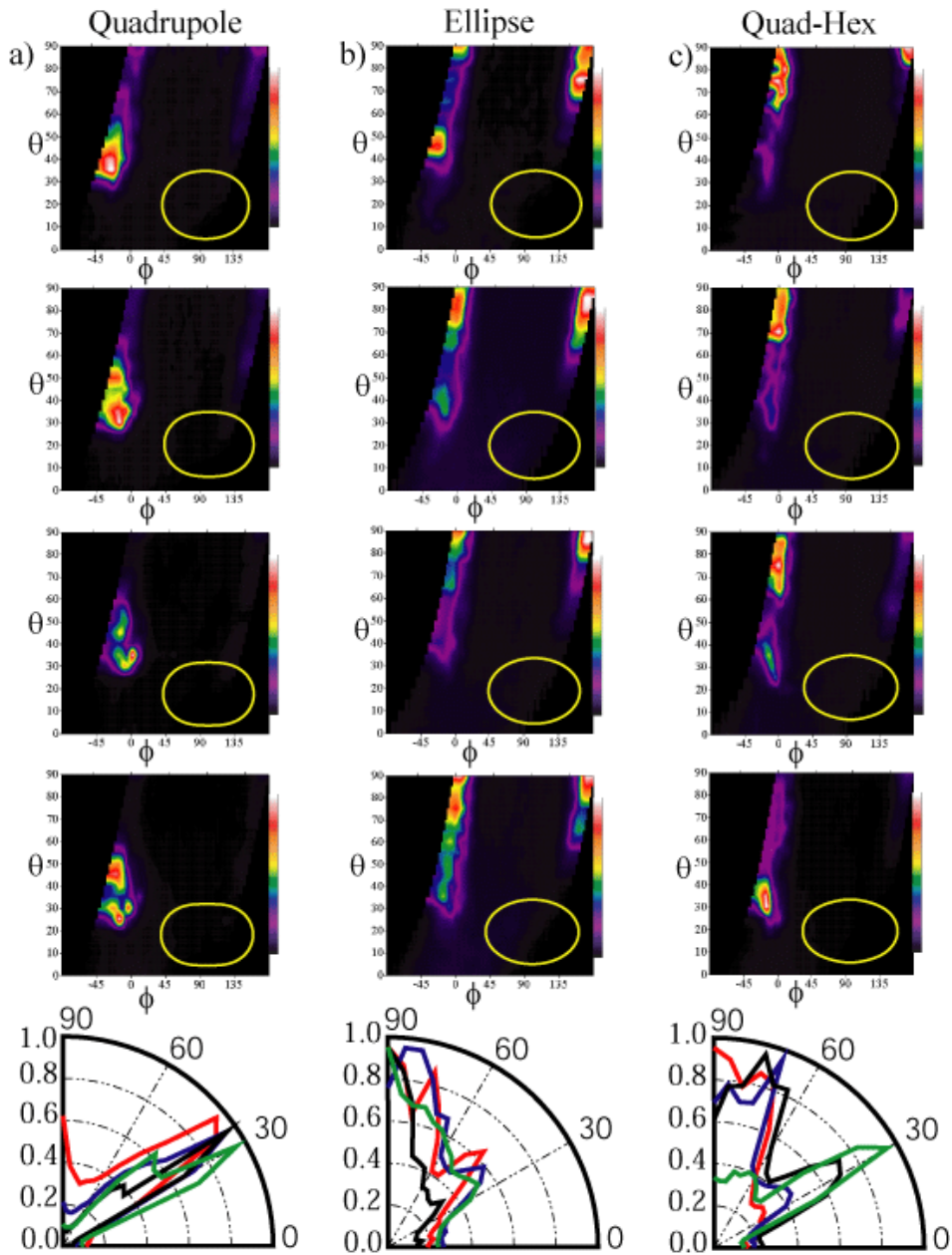}
\caption{Two-dimensional display of the experimental data showing in false color scale the emission intensity as a function of sidewall angle $\phi$ (converted from ICCD images) and of the far-field angle $\theta$ (camera angle).  Columns from left to right represent the quadrupole, ellipse and quadrupole-hexadecapole respectively. Insets show the cross-sectional shapes of the pillars in each case (for definitions see Fig.~\ref{fig:shapes}). The graphs at the bottom show the far-field patterns obtained by integration over $\phi$ for each $\theta$, normalized to unity in the direction of maximal intensity. The deformations are $\varepsilon=0.12, 0.16, 0.18, 0.20$ (red, blue, black and green respectively)}
\label{fig:ExpData}
\end{figure*}

The experiments we report were performed on differently shaped dye (DCM)-doped polymer (PMMA) samples that are fabricated on top of a spin-on-glass buffer layer coated over a silicon substrate via a sequence of microlithography and O$_2$ reactive ionic etching steps. The effective index of refraction of these microcavities is $1.49$, much lower than for earlier experiments performed using a similar set-up on GaN, where the index of refraction is $n=2.65$\cite{rex99,rex02}. They are optically pumped by a Q-switched Nd:YAG laser at $\lambda=532$ nm incident normal to the plane of the micropillar. Light emitted from the laser is imaged through an aperture subtending a $5^\circ$ angle and lens  onto a ICCD camera which is rotated by an angle $\theta$ in the far-field from the major axis. A bandpass filter restricts the imaged light to the stimulated emission region of the spectrum. The ICCD camera records an image of the intensity profile on the sidewall of the pillar as viewed from the angle $\phi$ which is converted from pixels to angular position $\phi$. In this paper we studied micro-cavities with elliptic, quadrupolar and quadrupolar-hexadecapolar shape of an average radius $R_o=100\mu$m (see formulas in Fig.~\ref{fig:shapes} caption). Each shape was analyzed at eccentricities of $\varepsilon=0.12, 0.14, 0.16, 0.18 \mbox{ and } 0.20$.

In Fig.~\ref{fig:ExpData} we show the experimental results in a color scaled ICCD image. The two angles are the sidewall angle $\phi$ (for the horizontal coordinate) and the camera angle $\theta$ (for the vertical coordinate). We omit the data for $\varepsilon=0.14$ deformation as it indicates no effects not captured by the data at the other deformations. To obtain the {\em far-field} pattern with respect to $\theta$ we integrate the image over all sidewall angles $\phi$. The {\em boundary image field} is calculated by integrating over $\theta$ for each $\phi$. As insets we show the exact shape of each of the microcavities. Although the shapes appear very similar to the eye, we find dramatic differences in the far-field emission patterns, which in the case of the ellipse vs.\ the quadrupole, persist over a wide range of deformations. Specifically, the far-field emission intensity for the quadrupole exhibits a strong peak at $\theta=34^{\circ}-40^{\circ}$ which remains rather narrow over the observed range of deformations. Over the same range of deformation the boundary image field for the quadrupole changes substantially and does not exhibit one localized point of emission. In contrast, the ellipse emits into the $\theta \sim 90^\circ$ direction in the far-field, but with a much broader angular intensity distribution, while the boundary image field remains well-localized around $\phi \sim 0^\circ$ (the point of highest curvature in the imaged field). For the quadrupole-hexadecapole we see a far-field directionality peak which shifts from $\theta\sim 90^\circ$ to $\theta \sim 30^\circ$ and an almost constant boundary image field. Thus we see three qualitatively different behaviors for the three shapes studied over the same range of variation of the major to minor axis ratios.

Several different samples with the same boundary shape were measured in each case and confirmed that the basic features of this data set just described are reproduced within each class (with small fluctuations)\cite{rex_thesis}. This shows that the effects measured are a property of the boundary shape and not of uncontrollable aspects of the fabricationprocess. Moreover the theoretical calculations, which we will present next, are based on uniform dielectric rods with the ideal cross-sectional shape specified by the mask; therefore the agreement of these calculations with the measurements also confirms that the differences are due to controllable shape differences.

\section{Theoretical Calculations}
\label{section:theo}
\begin{figure}[hbt]
\centering
\includegraphics[width=5cm]{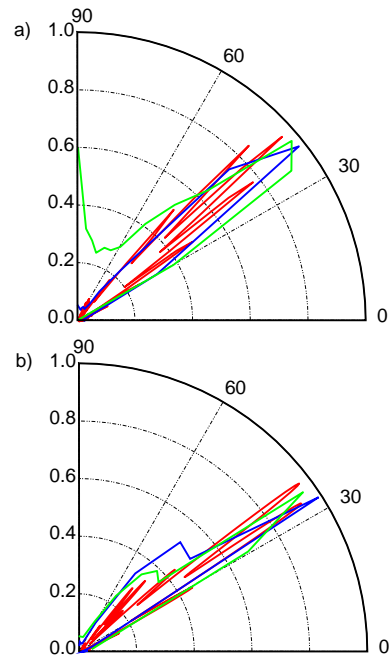}
\caption{Far-field intensity for the quadrupole with $\varepsilon=0.12$ (a) and $0.18$ (b). The green curve is the experimental result, blue the ray simulation and red a numerical solution of the wave equation. The ray simulation was performed starting with 6000 random initial conditions above the critical line and then propagated into the far-field in the manner described in the text. The numerical solutions selected have $kR_0=49.0847 -0.0379i$ with a $Q=-2\RE[kR_0]/\IM[kR_0]=2593.05$ and $kR_0=49.5927  -0.0679i $ with $Q=-2\RE[kR_0]/\IM[kR_0]=1460.72$ for $\varepsilon=0.12$ and $0.18$ respectively.}
\label{fig:FFquad}
\end{figure}
\begin{figure}[hbt]
\centering
\includegraphics[width=5cm]{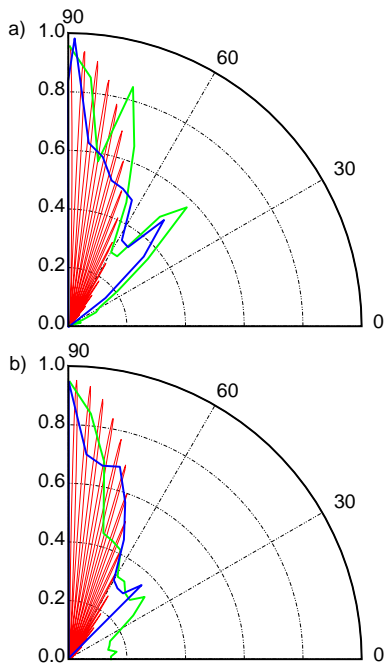}
\caption{Far-field intensity for the ellipse with $\varepsilon=0.12$ (a) and $0.18$ (b). Green, blue and red curves are experiment, ray simulation and wave solution. The ray simulation was performed starting with 6000 initial conditions spread over seven caustics separated by $\Delta\sin\chi=0.02$ below the critical caustic (the caustic that just touches the critical line). The numerical wave solutions shown correspond to $kR_0=49.1787 -0.0028i$ with $Q=-2\RE[kR_0]/\IM[kR_0]=17481.38$ and $kR_0=49.2491 -0.0110i$ with $Q=-2\RE[kR_0]/\IM[kR_0]=4488.20$ for $\varepsilon=0.12$ and $0.18$ respectively.}
\label{fig:FFellipse}
\end{figure}
\begin{figure}[hbt]
\centering
\includegraphics[width=5cm]{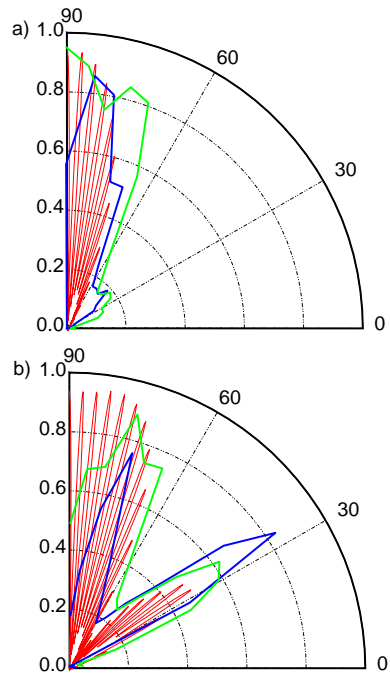}
\caption{Far-field intensity for the quadrupole-hexadecapole with $\varepsilon=0.12$ (a) and $0.18$ (b). Green, blue and red curves are experiment, ray simulation and wave solution as described in the caption to Fig.~\ref{fig:FFquad}. The numerical wave solutions shown correspond to $kR_0=50.5761  -0.0024i$ with $Q=-2\RE[kR_0]/\IM[kR_0]=42573.77$ and $kR_0=49.5642 -0.0092i$ with $Q=-2\RE[kR_0]/\IM[kR_0]=10741.93$ for $\varepsilon=0.12$ and $0.18$ respectively.}
\label{fig:FFoktu}
\end{figure}

In Figs.~\ref{fig:FFquad}-\ref{fig:FFoktu} we compare the experimental results for the far-field emission patterns for the three shapes measured at $\varepsilon = 0.12, 0.18$ to two theoretical models, one based on solutions of the wave equation and the other based on simulations of ray escape. The agreement in both cases is quite good. We briefly summarize here the two models used.

It is conventional in treating high-Q laser cavities to approximate the lasing mode by the appropriate resonant mode of the passive cavity (the cavity in the absence of pumping and gain)\cite{siegman_book}. In the current case we can model the passive cavity as a 2-d resonator because the penetration depth of the pump laser is of the order of $\mu$m and amplification due to reflection from the sidewall occurs primarily within the horizontal plane of the resonator. The wave-equation for the time-harmonic solutions in 2-d within the uniform dielectric reduces to the Helmholtz equation for the electric field (TM modes) and magnetic field (TE modes) and we can solve the Helmholtz equation by imposing Sommerfeld boundary conditions (no incoming wave from infinity). We use an efficient new numerical method developed for these systems described in ref.\cite{TureciSchwefel03} The method yields a large set of resonances over the entire range of Q-values; the real part of the wavevector giving the resonance frequency and the imaginary part giving its width. Experiments have indicated that mode selection is complicated in these dielectric resonators and that there is no simple rule relating the observed lasing mode to the Q-value of the mode in the passive cavity. Due to the low output coupling high Q-modes are not necessarily the observed lasing modes in the far-field. Thus from the set of calculated resonances we choose the resonance which coincides well with the observed far-field pattern and has a relatively high $Q=-2\RE[k]/\IM[k]$. We also confirmed that theoretical boundary image data\cite{rex02} coincides well with the experimental results. Moreover in all cases discussed here, there were many resonances which gave good agreement with the data, indicating the existence of a robust class of modes any of which could be the lasing mode. In Fig.~\ref{fig:FFquad}-\ref{fig:FFoktu} we as red line the numerical far-field by calculating the asymptotic expansion of our wavefunction in the far-field. Numerical limitations prevent us from performing the calculations at the experimental values of $kR_0 \sim 1000$ but the major features of the emission pattern are not sensitive to $kR_0$ over the range we can study numerically. The finding (discussed next) that we can reproduce these patterns from ray escape simulations also suggests that the wavelength is not a relevant parameter for the features we are studying. With a green line we show the experimental results. We see that the numerical calculations agree with the measured far-field very well.

As just noted, the experiment is performed well into the short-wavelength limit, and we are motivated to develop ray-optical models for the emission, following references\cite{NockelSCGC96,NockelS97}. In the ray simulation an initial distribution of rays is assumed and each ray is given unit initial amplitude. At each reflection on the boundary the amplitude is reduced according to Fresnel formulas. The outgoing amplitude is recorded in the direction determined by Snell's law and the reflected ray is followed until its amplitude falls below $10^{-4}$. To compare to the experimental data we collected the transmitted rays in $5^\circ$ bins. A subtle issue in the calculations is the choice of the initial ray distribution. In Figs.~\ref{fig:FFquad} and~\ref{fig:FFoktu} we show the far-field distribution for a randomly chosen set of initial conditions above the critical angle; in the case of the ellipse (Fig.~\ref{fig:FFellipse}) we chose initial conditions appropriate to its integrable dynamics (see the discussion below). The ray model is found to reproduce the main features of the data quite well. In section~\ref{section:adiab} we show that the far-field emission pattern for chaotic shapes is insensitive to the initial ray distribution over a wide range. Specifically, in Fig.~\ref{fig:FFquadInitial} below we compare the far-field patterns for different possible initial distributions, confirming the approximate independence of the patterns to this choice.

\section{Novel features of the Data}
\label{section:surpise}
In the previous section we showed that we can reproduce the experimental data with two different theoretical models. First, by solving the linear wave equation in the dielectric and choosing an appropriate resonance; second, by ray escape simulations. This gives us confidence that the major differences in the experimental emission patterns are due to the different shapes of the laser cavities. The strong sensitivity of the emission patterns to small differences in boundary shape is quite striking and is a major result of this study. This sensitivity was predicted in the earlier work of references\cite{NockelSCGC96,NockelS97,ChangCSN00} and was not unexpected. However there are major aspects of the experimental data which are quite surprising even in the light of the earlier work on ARCs. In particular, the persistence of highly directional emission in the quadrupolar shapes at quite high deformations was not predicted theoretically and was unexpected for reasons we will now discuss. In order to understand the unexpected features of the data and to develop principles to predict the emission patterns for untested boundary shapes we will review and extend the phase space approach introduced originally by N\"ockel and Stone.

To explain the basic concepts it is useful to begin by neglecting the possibility of refraction out of the resonator. The problem of ray motion in a resonator with perfectly reflecting walls is equivalent to the billiard problem first posed by Birkhoff\cite{birkhoff27} and since widely studied in the non-linear dynamics community.  The billiard problem is that of a point mass moving freely in two dimensions within a boundary of general shape defined by perfectly (specular) reflecting wall. In such a system classical mechanics is reduced to a simple geometric construction, whose properties nonetheless turn out to be remarkably complex. An important tool to study dynamical systems such as billiards was devised by Poincar\'e\cite{poincare1892}, the {\em Poincar\'e Surface of Section} (SOS). The SOS images the trajectories in phase space instead of in real space. For even the simplest non-trivial dynamical systems (two dimensions, one degree of freedom) the phase space is four-dimensional, however the SOS is a two-dimensional plot which is a stroboscopic image of a set of trajectories as they cross a certain hyper-plane in phase space. For billiards this image is taken at each reflection from the boundary; the coordinates of the SOS are the angular position ($\phi$) and the tangential momentum $\sin \chi$ (scaled to unit total momentum) of each trajectory as it reflects from the boundary (Fig.~\ref{fig:shapes}). The topology of the resulting plot gives us critical information about the phase space flow in the particular billiard of interest.

\begin{figure}[htb]
\centering
\includegraphics[width=\linewidth]{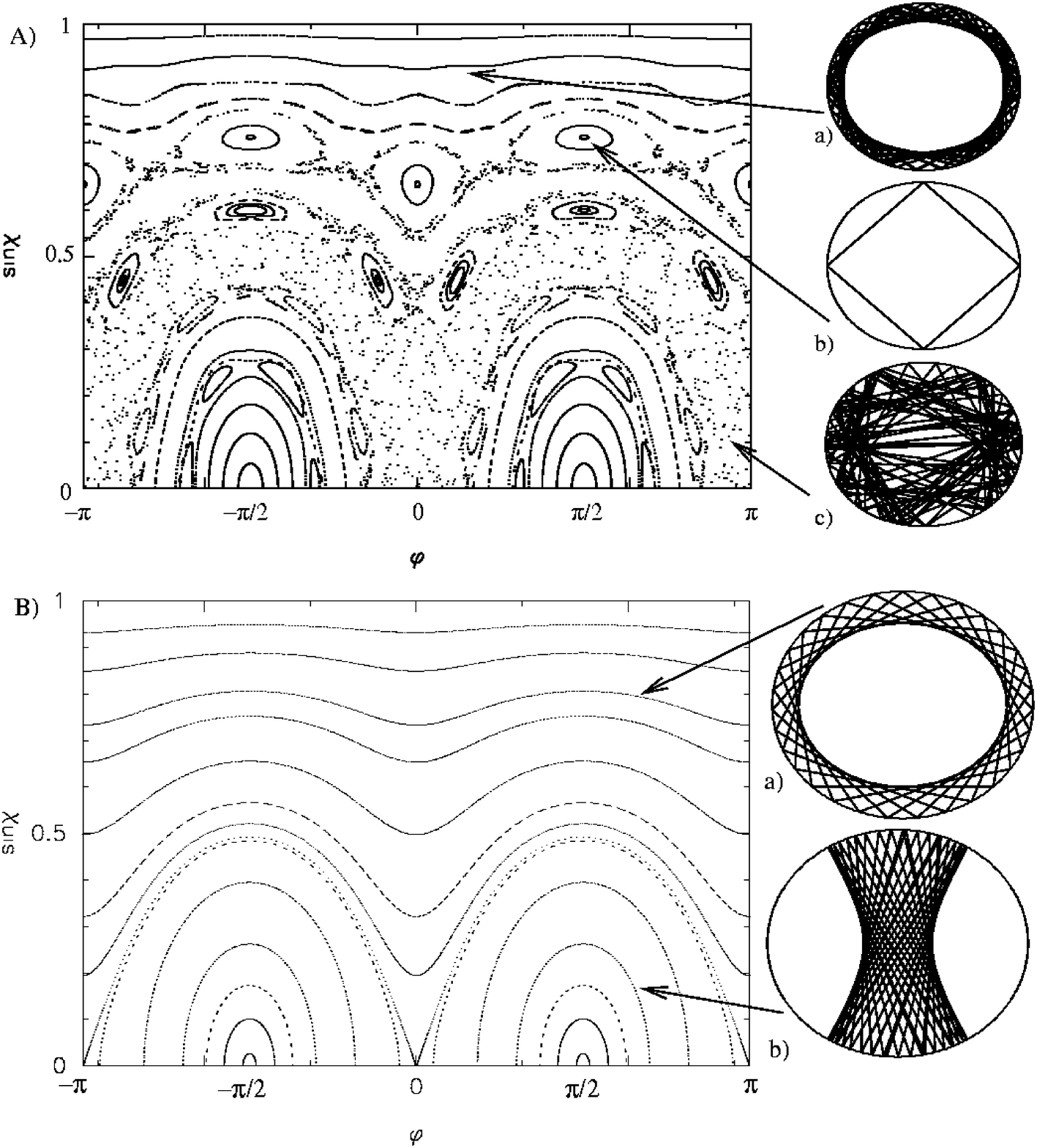}
\caption{The Poincar\'e surface of section for the quadrupole A) and the ellipse B) with $\varepsilon=0.072$. The schematics A)(a-c) on right show three classes of orbits for the quadrupole, A)(a) a quasi-periodic orbit on a KAM curve, A)(b) a stable period-four orbit, (the `diamond'), and A)(c) a chaotic orbit. Schematic B)(a, b) show the two types of orbits which exist in the ellipse, the whispering gallery type, with an elliptical caustic B)(a) and B)(b), the bouncing ball type, with a hyperbolic caustic.}
\label{fig:SOS0.072}
\end{figure}

When the motion is partially chaotic a real-space trace of the trajectory is essentially useless, while the SOS can reveal useful information. An example of this is shown in Fig.~\ref{fig:SOS0.072}A). Here we show a SOS for the quadrupole and the ellipse with a deformation of $\varepsilon=0.072$; we calculated and plotted trajectories arising from $\sim30$ different initial conditions and iterated them for 500 reflections. We see that for the chaotic orbit Fig.~\ref{fig:SOS0.072}A)(c) the real space picture does not provide any useful information, whereas the SOS shows us an underlying structure; specifically that such a chaotic orbit nonetheless avoids large regions of phase space in this case.  This plot illustrates a typical structure for a {\em mixed phase space}. Such a structure is generic for all smooth deformations of a circular billiard although the fraction of phase space which is covered by chaotic motion increases with deformation. We will not attempt to review the general properties of mixed dynamical systems which are described by
Kolmogorov-Arnold-Moser (KAM)\cite{arnold_book} theory and by other theorems due to Poincar\'e, Birkhoff and Lazutkin. However a key to understanding a given billiard is to study its periodic orbits.  The ``diamond" periodic orbit in Fig.~\ref{fig:SOS0.072}A)(b) shows up in the SOS as four points which are referred to as fixed points in the billiard map. For this deformation the diamond orbit is called {\em stable} as motion close to the fixed points is bounded in the vicinity of the fixed points. Iteration of a point close to a stable fixed point will result in a 1-d set of points and is called {\em quasi-periodic orbit} as it fills a 1-d curve densely. The set of all the points in the neighborhood of a stable fixed point that are quasi-periodic is called an {\em island}. Another type of quasi-periodic motion is shown in Fig.~\ref{fig:SOS0.072}A)(a); in this case we have motion on a ``KAM" curve which is similar to the WG modes found in the circle and is not associated with any stable periodic orbit. The final type of motion which occurs in a mixed phase space is shown in Fig.~\ref{fig:SOS0.072}A)(c). Here we show a chaotic orbit which fills a {\em two-dimensional} region of the SOS densely; hence this is a qualitatively different type of motion. Within this region nearby initial conditions will separate exponentially in time.(until they are separated by a distance of order the size of teh chaotic region). As noted above, the existence of these three types of motion is generic for smooth deformations of a circle and hence for ARC resonators.

While the behavior of the quadrupole shown in Fig.~\ref{fig:SOS0.072}A) is generic there do exist special billiards that exhibit the two extremes of dynamical behavior. One limiting case is exemplified by the ellipse billiard whose SOS is shown in Fig.~\ref{fig:SOS0.072}B). For the ellipse all orbits lie on 1-d curves in the SOS and there are no chaotic regions of phase space. This type of system is called integrable and has regular dynamics because there exists one constant of motion for each degree of freedom. For the elliptical billiard these constants are the energy and the product of the angular momenta with respect to each focus (degenerating to the angular momentum in the case of the circular billiard)\cite{berry81a,noeckel_thesis}.  The ellipse is the only convex deformation of a circular billiard which is integrable\cite{poritsky50}. A recent proof of this long standing conjecture was given by Amiran\cite{amiran97}. At the opposite extreme is the Bunimovich stadium billiard (see inset in Fig.~\ref{fig:FFstadium}) for which it is proven that there exist no stable periodic orbits and the entire phase space (except sets of measure zero) is chaotic. We will study theoretically the emission from stadium-shaped resonators in section~\ref{sect:stadium}.

Due to its integrability, phase space flow in the ellipse is particularly simple: every initial condition lies on one of the invariant curves given by Eq.~(\ref{eq:adiab}) below, and the trajectory retraces this curve indefinitely (see Fig.~\ref{fig:SOS0.072}B)).  Curves which cross the entire SOS correspond to real-space motion tangent to a confocal elliptical caustic Fig.~\ref{fig:SOS0.072}B)(a); curves which do not cross the entire SOS represent motion tangent to a hyperbolic caustic in real space Fig.~\ref{fig:SOS0.072}B)(b).

Phase space flow in mixed systems is much more complex and is ergodic on each chaotic region. However a key property of mixed dynamical systems is that the different dynamical structures in phase space are disjoint; this implies that in two dimensions KAM curves and islands divide phase space into regions which cannot be connected by the chaotic orbits.  This puts constraints on phase space flow despite the existence of chaos in a significant fraction of the phase space. For small deformations ($\sim 5$\%) most of phase space is covered by KAM curves, the form of which can be estimated using an adiabatic approximation\cite{NockelS97}. This approximation gives the exact result for all deformations in the case of the ellipse; it can be written in the following form:
\be
\sin\chi(\phi)=\sqrt{1+(S^2-1)\kappa^{2/3}(\phi,\varepsilon)}
\label{eq:adiab}
\ee
where $\kappa$ is the radius of curvature along the boundary and $S$ is a constant. Plotting this equation for different values of $S,\varepsilon$ gives an SOS of the type shown in Fig.~\ref{fig:SOS0.072}B). For the mixed case, exemplified by the quadrupole billiard in Fig.~\ref{fig:SOS0.072}A), Eq.~(\ref{eq:adiab}) describes quite accurately the behavior for values of $\sin \chi $ near unity, but doesn't work well at lower $\sin \chi$ where chaos is more prevalent.

Up to now the discussion has been based on the ideal billiard, where a ray is trapped indefinitely by perfectly reflecting walls. N\"ockel and Stone realized that in this short wavelength limit the dielectric resonator would behave very similar to the ideal billiard for angles of incidence above $\sin \chi_c = 1/n$ (i.e.\ the evanescent leakage could be neglected), however for incidence below $\sin \chi_c$ rays would rapidly escape by refraction according to Fresnel's law. If a correspondence could be made between a set of initial conditions for rays and a set of solutions of the wave equations (modes) then the emission pattern could be calculated by propagating those rays forward in time and allowing them to escape into the far-field according to Snell's and Fresnel's laws. For the case of the ellipse there was an obvious correspondence between a set of rays chosen on a given invariant curve of the ellipse and a set of solutions. This correspondence can be formalized using the eikonal method of Keller\cite{Keller60}. The difficulty was in finding the correspondence in the chaotic regions of phase space of generic shapes, for which there exist no invariant curves.
\begin{figure}[t]
\centering
\includegraphics[width=0.8\linewidth]{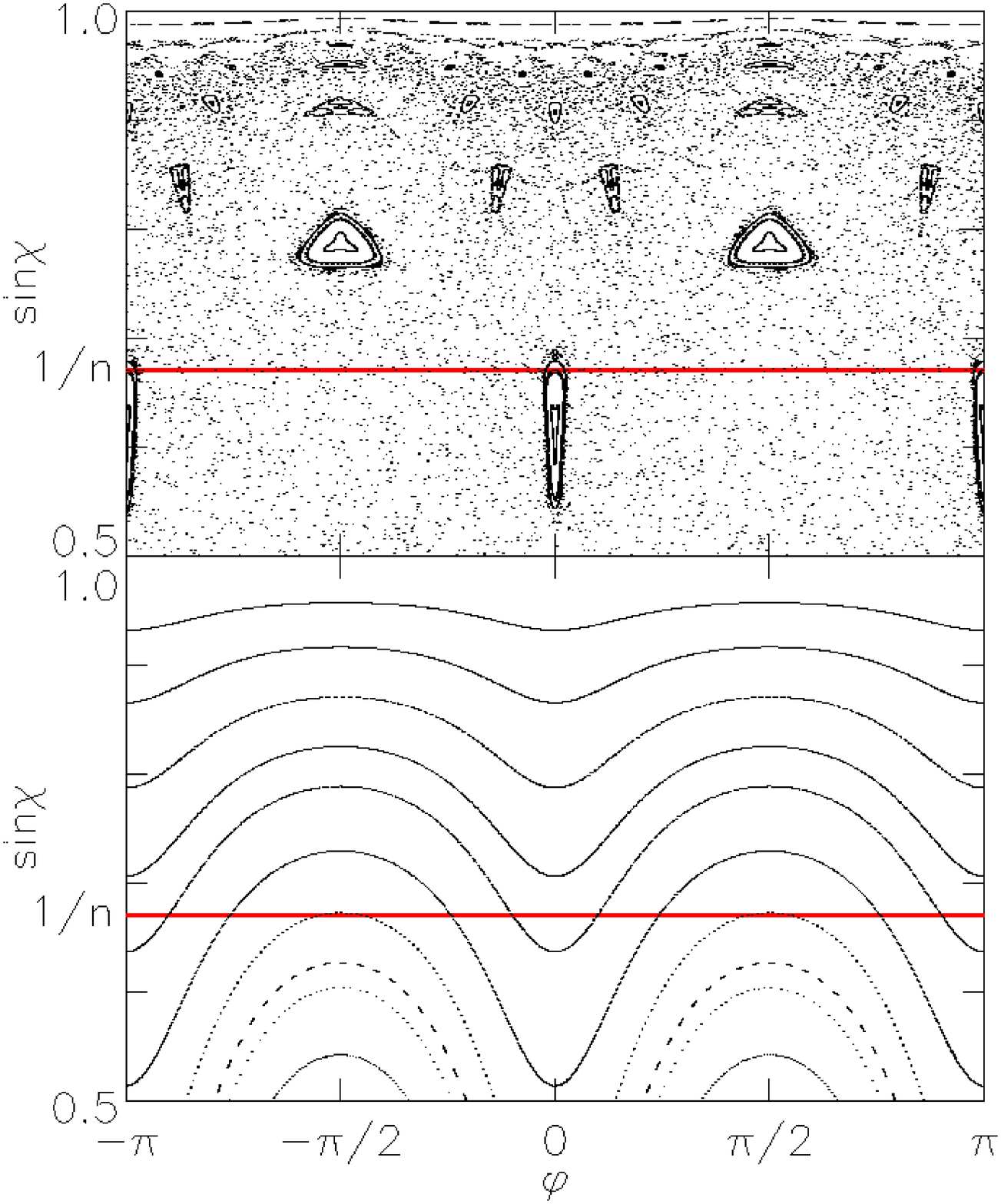}
\caption{Comparison of the Poincar\'e surface of section for the quadrupole and the ellipse with $\varepsilon=0.12$ showing mostly chaotic behavior in the former case and completely regular motion in the latter. The red line denotes $\sin\chi_c=1/n$, the critical value for total internal reflection; rays above that line are trapped and those below escape rapidly by refraction. The quadrupole still exhibits stable islands at $\phi=0, \pi$ and $\sin\chi=\sin\chi_c$ which prevent escape at the points of highest curvature in the tangent direction}
\label{fig:SOS0.12}
\end{figure}

In refs.\cite{noeckel_thesis,NockelS97} a model was proposed based on an adiabatic theory to describe the ray-wave correspondence in the generic case of mixed dynamics. Initial conditions on the adiabatic invariant curves, e.g.\ of the quadrupole were chosen. Due to the presence of chaos in the true dynamics, rays initially on such a curve would diffuse in phase space until they escaped by refraction. The resulting emission pattern can be calculated by ray simulations of the type we have presented above. Moreover this model led to qualitative predictions about the emission patterns without doing any simulations. The adiabatic invariant curves for the quadrupole have their minimum values of $\sin\chi$ at the points of highest curvature on the boundary $\phi=0,\pm\pi$, just as they do in the ellipse. If the diffusion in phase space is sufficiently slow, emission would be near these points of highest curvature and at the critical angle, i.e.\ in the tangent direction, as in the ellipse. This reasoning held as long as the escape points $\sin\chi=1/n, \phi=0,\pm\pi$ occurred in the chaotic region and were reachable from the totally-internally-reflected region of $\sin\chi>1/n$.  However for $n=1.5$ and deformations around $10$\%, these points are enclosed by the stable island corresponding to the four-bounce ``diamond'' orbit and due to the disjoint nature of the dynamics, ``chaotic'' rays cannot escape there.  Instead they will escape at higher or lower values of $\phi$ leading to a large change in the emission pattern from that of the ellipse with similar minor-major axis ratio. This is the phenomenon termed ``dynamical eclipsing'', and it was predicted to occur for the $n=1.5$ quadrupole at $\varepsilon \sim 0.12$ some time ago\cite{NockelS97,NockelSCGC96}. Our experimental data confirms this prediction for the $\varepsilon=0.12$ case. Fig.~\ref{fig:SOS0.12} contrasts the phase space for the ellipse and the quadrupole for this deformation. The island associated with the stable diamond orbit is smaller than at $\varepsilon=0.072$, but still present for the quadrupole; there is no such island at any deformation for the ellipse. Note that in the experimental data for the quadrupole at $\varepsilon = 0.12$ we do not see a bright spot at the boundary at $\phi=0$, consistent with the dynamical eclipsing model in which the island structure forces the chaotic WG modes to emit away from the point of highest curvature.  In contrast the bright spot in the ellipse which emits to $\theta=90^{\circ}$ clearly is at $\phi=0$ for $\varepsilon = 0.12$.  Thus the adiabatic model of refs.\cite{NockelS97,NockelSCGC96} does seem to provide a reasonable description of the data for $\varepsilon = 0.12$ and the observed dramatic difference between the ellipse and quadrupole shapes is as predicted.

\begin{figure}[htb]
\psfrag{epsilon}{$\varepsilon$}
\psfrag{trace}{$\scriptstyle TrM$}
\centering
\includegraphics[width=\linewidth]{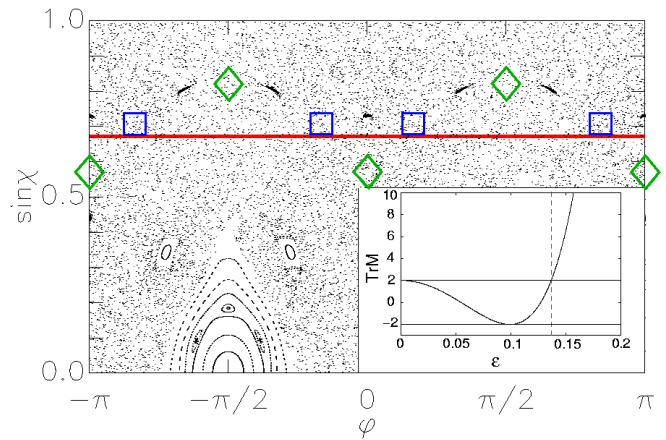}
\caption{Poincar\'e surface of section for the quadrupole with $\varepsilon=0.18$. The red line indicates the critical angle of incidence. The green diamonds indicate the location of the fixed points of the (now) unstable ``diamond'' orbit and the blue squares the fixed points of the unstable rectangular orbit. In the inset we show the trace of the monodromy (stability) matrix (see Eq.~\ref{eq:linMap}) for the diamond orbit versus deformation. When the magnitude of the trace of the monodromy matrix is larger than two its eigenvalues become real, the periodic motion becomes unstable and the associated islands vanish.  For the diamond this happens at $\varepsilon = 0.1369$ (see dashed vertical line in the inset) and the simple dynamical eclipsing picture of Fig.~\ref{fig:SOS0.12} does not apply at larger deformations.}
\label{fig:SOS0.18}
\end{figure}

The earlier work on ARCs did not look extensively at deformations above $\varepsilon = 0.12$ for the case of low index materials such as polymers.  The belief was that the adiabatic model would become questionable at higher deformations as the phase space became more chaotic and the ray motion departed from the adiabatic curves very rapidly. A natural expectation was that due to increased chaos the emission patterns in the far-field would become less directional and more pseudo-random.  More specifically, for the $n=1.5$ quadrupole one finds that the ``diamond'' orbit becomes unstable at $\varepsilon \approx 0.1369$ (the associated islands shrink to zero) and one would not expect highly directional emission at higher deformations (see inset in Fig.~\ref{fig:SOS0.18}). Thus a plausible extrapolation of the adiabatic model suggests a steady broadening of the quadrupole emission with deformation with at least some significant emission in the tangent ($90^{\circ}$) direction.  Our experimental data strongly contradicts this expectation, as the observed emission patterns remain peaked around $35^{\circ}$ and do not broaden at all up to $\varepsilon=0.20$. A similar analysis would show that the adiabatic theory provides no qualitative explanation for the switch in the far-field directionality for the quadrupole-hexadecapole at high deformations. Thus we are motivated to look for a model of the phase space flow which can explain the persistence in highly directional emission at high deformations where almost all of the phase space is chaotic. This model will be presented in the next section.

\section{Short-time dynamics and Unstable Manifolds}
\label{section:adiab}
\begin{figure}[thb]
\centering
\includegraphics[angle=-90,width=0.9\linewidth]{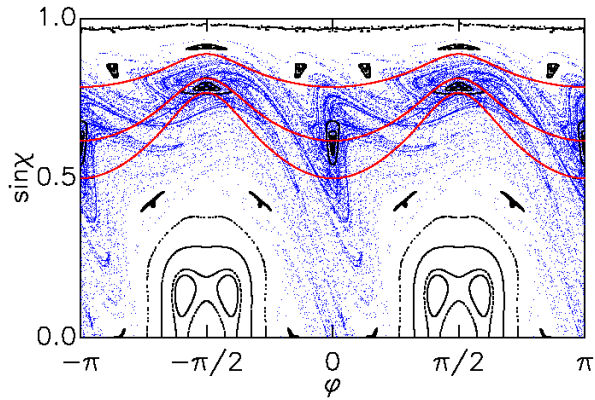}
\caption{Simulation of short-term ray dynamics for the quadrupole with deformation $\varepsilon=0.12$, shown in blue. The three red curves are examples of adiabatic curves for the quadrupole; the blue points are the result of iterating 8,000 starting rays for 50 bounces starting on the middle of the three red curves. The blue points show a complex structure which is asymmetric around the line $\phi=0$ in contrast to the SOS and the adiabatic curves which have reflection symmetry.}
\label{fig:adiab0.12}
\end{figure}

At high deformations chaotic diffusion is fast and rays tend to escape rapidly even if they are initially well confined (i.e.\ far away from the critical angle). The adiabatic model reviewed predicted at which regions on the boundary rays are more likely to escape.  How accurate is this model as a description of the phase space flow?  One easy thing to check is whether the short-term flow preserves the discrete symmetries of the system; the shape of the adiabatic curves are determined by the shape of the boundary, via its curvature (see Eq.~\ref{eq:adiab}) and thus must preserve all the billiard symmetries (e.g.\ horizontal and vertical reflection axes for the quadrupole).  A straightforward ray simulation shown in Fig.~\ref{fig:adiab0.12} reveals that the short-term dynamics typically breaks these symmetries.  We begin a set of rays on an adiabatic curve, which has reflection symmetry around $\phi=0$, and after 50 iterations the SOS density has developed a strong asymmetry around the origin.  Thus the flow cannot be accurately seen as simple diffusion between adiabatic curves. On the other hand the resulting SOS density is very structured and does not look like a realization of independent random walks.  We shall see below that the boundaries of the high density regions are set by the unstable manifolds of a few short periodic orbits.

\begin{figure}[hbt]
\centering
\includegraphics[width=5cm]{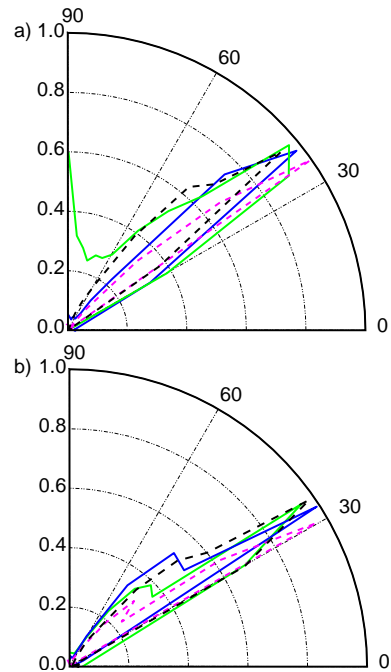}
\caption{Ray simulations of the far-field emission patterns for the quadrupole with $\varepsilon=0.12$ (a), $ \varepsilon = 0.18$ (b) with different types of initial conditions. The blue curve is the result of choosing random initial conditions about the critical line $\sin \chi = 1/n$, the black dashed curve is for initial conditions on the adiabatic curve with minimum value at the critical line. The magenta curve is for initial conditions localized around the unstable fixed point of the rectangle periodic orbit. In each of the ray simulations 6000 rays were started with unit amplitude and the amplitude was reduced according to Fresnel's law upon each reflection, with the refracted amplitude ``collected'' in the far-field. The green curve is the experimental result; clearly all three choices give similar results in good agreement with experiment.}
\label{fig:FFquadInitial}
\end{figure}

Another indication that the emission patterns we observe are not closely linked to adiabatic curves is obtained by looking at the dependence of the ray emission directionality on initial conditions. The original work of N\"ockel and Stone began the ray simulations with an ensemble of rays uniformly distributed on the adiabatic curves. This is obviously correct for the ellipse as one can calculate resonances which cover uniformly the different invariant curves. However, as already noted, the ellipse is special since chaotic diffusion is absent. In Fig.~\ref{fig:FFquadInitial} we compare ray emission patterns arising from three different choices of initial conditions in the quadrupole: uniform on the adiabatic curve (black dashes), initial conditions localized on the unstable fixed points associated with the rectangular periodic orbit shown in Fig.~\ref{fig:shapes} (magenta dashes), and finally, initial conditions chosen {\it randomly} above the critical angle for escape (blue). These three quite different choices all lead to similar far-field intensity patterns, in good agreement with the experimental measurements (green). More generally, we found that the ray simulations are quite insensitive to the choice of initial conditions, as long as a significant fraction of the rays are started within the chaotic sea. (Ray bundles only started in an island would obviously lead to different results.)  We shall propose an explanation for this insensitivity in the next section.

\section{Unstable Manifolds}
\label{section:manifold}
By definition a non-linear system which is partially chaotic, such as our billiards, generates ergodic and mixing motion on a finite fraction of the phase space for long times.  It is convenient to analyze the dynamics in the SOS and not in continuous time; in this case the dynamics is described by a discrete map. One can get a good idea of the short term dynamics of such a map in selected regions of the SOS by linearizing it in the neighborhood of unstable fixed points of the billiard map (corresponding to unstable periodic orbits in real space). If we take the initial position and direction/momentum of one ray at the boundary to be $(s,u)=(\phi, \sin\chi)$ we define the map which projects the ray to the next position and direction to be
\be
T:(\phi, \sin\chi)\rightarrow(\phi', \sin\chi').
\label{eq:map}
\ee
A set of fixed points of order N is defined by
\be
T^N(\phi, \sin\chi)=(\phi, \sin\chi).
\label{eq:fixpoint}
\ee

We can propagate an initial ray corresponding to a small deviation from the fixed point values by linearizing the map around the fixed point.
\be
T(\phi, \sin\chi)\sim M(\phi, \sin\chi)^T=\mat{\pd{s'(s,u)}{s}}{\pd{u'(s,u)}{s}}
{\pd{s'(s,u)}{u}}{\pd{u'(s,u)}{u}}(\phi, \sin\chi)^T
\label{eq:linMap}
\ee
The nature of the nearby motion can then be characterized by calculating the eigenvalues and eigenvectors of $M$. For Hamiltonian flows $M$ is always an area-preserving map, i.e.\ det$M=1$. The matrix $M$ is also known as the {\em monodromy}, or {\em stability} matrix. The eigenvalues can be either complex on the unit circle or purely real and reciprocal to each other. If the eigenvalues are complex, the fixed points are stable (elliptic) and nearby points oscillate around the fixed points tracing an ellipse in the SOS. In this case the long-time dynamics is determined by the linearized map to a good approximation. In the case of real eigenvalues there will be one eigenvalue with modulus larger than unity (unstable) and one with modulus less than unity (stable) and there will be two corresponding eigendirections (not usually orthogonal).  In the stable direction, deviations relax exponentially towards the fixed points; in the unstable direction deviations grow exponentially away from the fixed points.  Generic deviations will have at least some component along the unstable directions and will also flow out along the unstable direction. Therefore, in a short time generic deviations move out of the regime of validity of the linearized map and begin to move erratically in the chaotic ``sea''.  Hence the linearized map is not a good tool for predicting long time dynamics in a chaotic region of phase space.  However, in open billiards, rays will escape if they wander away from the fixed points into the part of the chaotic sea which is below the critical angle for total internal reflection. Therefore we find the unstable eigenvectors of the short periodic orbits useful in predicting ray escape.  For the short periodic orbits in our shapes it is possible to calculate the matrix $M$ giving the linearized map around all of the short periodic orbits. Thus we can calculate the eigendirections and determine the unstable directions analytically.

Before we can explain the characteristic asymmetric pattern found in Fig.~\ref{fig:adiab0.12}, we need to introduce the more general concept of the non-linear stable and unstable manifolds of the fixed points. Consider first the stable eigendirection defined by the the stable eigenvector of the monodromy matrix for a given fixed point. This is a line in the SOS passing through the fixed point such that any initial condition on that line will come arbitrarily close to the fixed point as $t \rightarrow \infty$; similarly the unstable direction is a line of initial conditions which approaches the fixed point arbitrarily closely as $t \rightarrow - \infty$.  At a certain distance in phase space from the fixed point the linearized map no longer holds and a deviation exactly on the line corresponding to the unstable eigenvector will {\it not} tend to the fixed point asymptotically; however nearby this line there exist such points which form a generalized curve known as the unstable {\it manifold} of the fixed point. Each unstable fixed point has associated with it stable and unstable manifolds which coincide with the eigendirections as one passes through the fixed point.  Note that for integrable systems there is only one asymptotic manifold for both past and future and it coincides exactly with the invariant curves, which can be calculated analytically in some cases (e.g.\ the ellipse).  For the non-integrable case, e.g.\ the quadrupole, we can only calculate the eigendirections near the fixed point analytically and must trace out the full manifolds numerically. As the unstable manifolds deviate further from the fixed points, generically they begin to have larger and larger oscillations.  This is necessary to preserve phase space area while at the same time have exponential growth of deviations. This tangling of the unstable manifolds has been used to devise a mathematical proof of chaotic motion\cite{smale67}.

One can argue qualitatively that the unstable manifolds of the short periodic orbits ought to control the ray escape dynamics at large deformations. The manifolds of short periodic orbits are the least convoluted as they are typically the least unstable; hence the unstable direction is fairly linear over a large region in the SOS. A typical ray will only make small excursions in phase space until it approaches one of these manifolds and then it will rather rapidly flow along it.  If the direction leads across the critical line for escape, that crossing point and the portion just below will be highly favored as escape points in phase space. We can check this qualitative argument with a few simple numerical experiments.
\begin{figure}[!thb]
\centering
\includegraphics[width=0.95\linewidth]{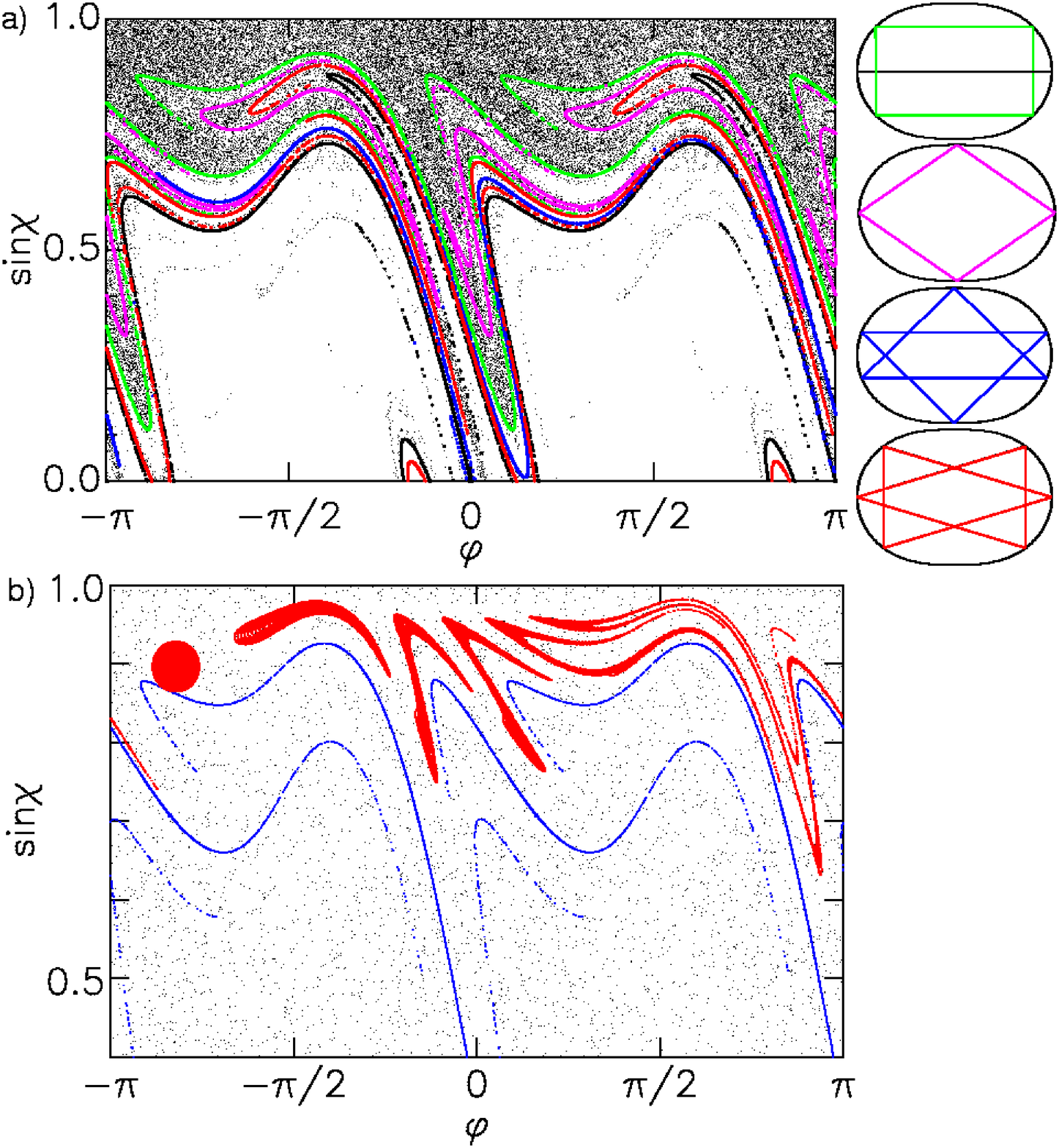}
\caption{(a) Ray simulations of short-term dynamics for random initial conditions above the critical line, propagated for 10 iterations, plotted on the surface of section for the quadrupole with $\varepsilon=0.18$. The areas of the SOS covered are delineated very accurately by the unstable manifolds of the short periodic orbits which are indicated in the schematics at right. These manifolds are overlaid in the figure with appropriate color coding. (b) Flow of phase space volume in the surface of section of the quadrupole with $\varepsilon=0.18$. A localized but arbitrary cloud of initial conditions (red) is iterated six times to illustrate the flow. The initial volume is the circle at the far left, successive iterations are increasingly stretched by the chaotic map. The stretching clearly follows closely the unstable manifold of the rectangle orbit which we have plotted in blue.}
\label{fig:ManifoldRandom0.18}
\end{figure}

In Fig.~\ref{fig:ManifoldRandom0.18}a)  we show the results of a short time iteration of a uniform random set of initial conditions above the critical line in comparison to the unstable manifolds of the various relevant short periodic orbits. Note that the different unstable manifolds must fit together in a consistent manner and cannot cross one another; if they did such a crossing point would define a ray which asymptotically in the past approaches two different sets of fixed points, which is not possible. Because of this non-crossing property the unstable manifolds define just a few major flow directions in the SOS. We see clearly in the simulation that the actual short-time flow of random points in the chaotic sea is controlled and approximately bounded by these unstable manifolds. To further support our statement that the general motion in phase space is governed by the unstable manifolds of these short orbits, in Fig.~\ref{fig:ManifoldRandom0.18}b)  we propagated an arbitrary but localized set of initial conditions and confirmed that they are stretched along and parallel to nearby unstable manifolds. Thus it appears that for the highly deformed case the phase space flow of a generic ray is much better predicted by simply plotting these manifolds.
\begin{figure}[thb]
\centering
\includegraphics[width=0.8\linewidth]{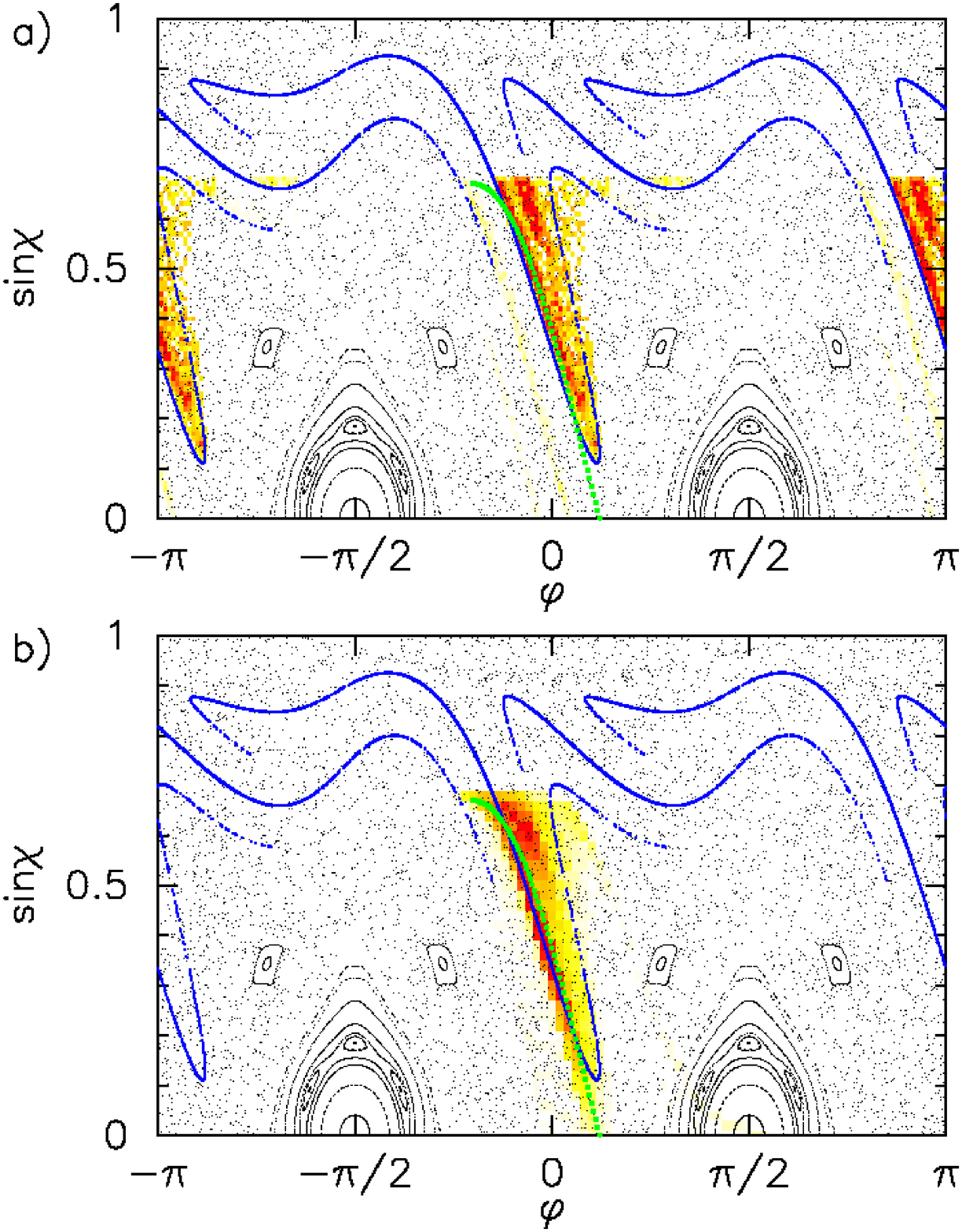}
\caption{a) Ray simulation of emission: emitted ray amplitude (color scale) overlaid on the surface of section for the quadrupole with $\varepsilon=0.18$. b) Far-field intensity from experimental image data Fig.~\ref{fig:ExpData} projected in false color scale onto the surface of section for the quadrupole with $\varepsilon=0.18$. The blue line is the unstable manifold of the periodic rectangle orbit. In green we have the line of constant $34^\circ$ far-field (see the discussion in section~\ref{section:constFarfield}). Absence of projected intensity near $\phi = \pm \pi$ in (b) is due to collection of experimental data only in the first quadrant.}
\label{fig:IntensitySOS}
\end{figure}

As a confirmation that these manifolds do control escape we perform a further ray simulation for the ``open" billiard. We propagate, as before, an ensemble of rays with a uniform random distribution above the critical angle.  As we have done in calculating the ray emission pattern, we associate to every starting ray in the surface of section an amplitude which decreases as the ray propagates forward in time according to Fresnel's law (if the point falls below the critical line).  Instead of following the refracted amplitude into the far-field, in this case we plot the {\it emitted} amplitude onto the surface of section, as shown in Fig.~\ref{fig:IntensitySOS}a). The emission amplitude is almost completely comfined within the two downwards ``fingers'' created by the unstable manifold of the four-bounce rectangular orbit. As noted earlier, the availabity of the two-dimensional data obtained from the imaging technique (see Fig.~\ref{fig:ExpData}), gives us a unique ability to reconstruct the emitting part of the lasing mode both in real space and momentum space directly from experimental data. It is therefore possible to check directly this ray simulation in phase space against experimental data. The intensity data is sorted into intensity pixels according to both its sidewall location (the angle $\phi$ from which emitted intensity originated) and its far-field angle, which by geometric considerations and Snell's law can be converted to the internal angle of incidence $\sin \chi$. Therefore we can project this data ``back" onto the SOS for emission. In Fig.~\ref{fig:IntensitySOS}b) we show this projection for the same deformation as in Fig.~\ref{fig:IntensitySOS}a); we find remarkable agreement between the projected data and the ray simulation. We note that this is a much more demanding test of agreement between theory and experiment than simply reproducing the experimental far-field patterns.

\section{Ray Dynamical Explanation of the Experimental Data}
\label{section:constFarfield}
\begin{figure}[thb]
\centering
\includegraphics[width=0.8\linewidth]{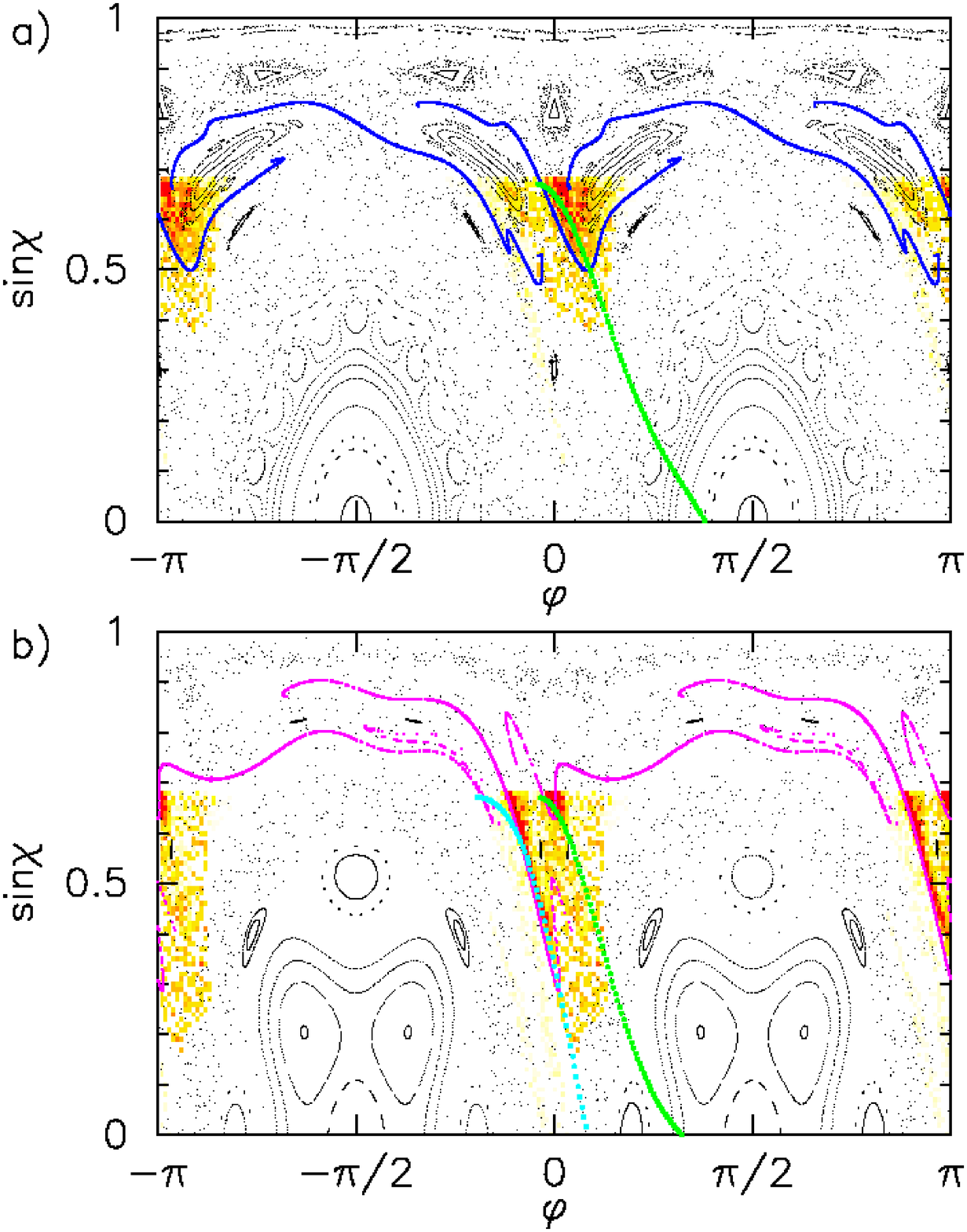}
\caption{Ray emission amplitude (color scale) overlaid on the surface of section for the quadrupole-hexadecapole with $\varepsilon=0.12$ (a) and $\varepsilon=0.20$ (b). Solid blue and magenta curves are the unstable manifold of the diamond orbit (a) and of the unstable rectangular orbit (b). In green and turquoise we plot the line of constant emission in the $75^\circ$ and $30^\circ$ directions in the far-field.}
\label{fig:IntensityOktu}
\end{figure}

In the last section we established that typical rays above the critical angle escape by following closely the unstable manifolds of the short periodic orbits. This leads to a ray escape probability which is relatively localized in the surface of section (Fig.~\ref{fig:IntensitySOS}a)).  However despite the non-random character of this escape, there is still a significant spread of angles of incidence for escape. In fact the spread of escape angle we see in Fig.~\ref{fig:IntensitySOS} would lead to an angular spread of nearly $80^\circ$ in the far-field if all the escape occurred from the same point on the boundary. However as we see from Fig.~\ref{fig:IntensitySOS}, the point of escape and the angle of incidence are correlated and vary together according to the shape of the unstable manifold.  Because the boundary is curved, different angles of incidence can lead to the same angular direction in the far-field. It is straightforward to calculate the curves of constant far-field for a given shape; for the quadruple at $\varepsilon=0.18$ and for the peak observed emission angle of $34^\circ$ this curve is plotted in green in Fig.~\ref{fig:IntensitySOS}. The curve tends to lie remarkably close to the unstable manifold.  Therefore we find that the curvature of the boundary tends to compensate almost completely for the dispersion in the angle of incidence at escape.
\begin{figure}[thb]
\centering
\includegraphics[width=5cm]{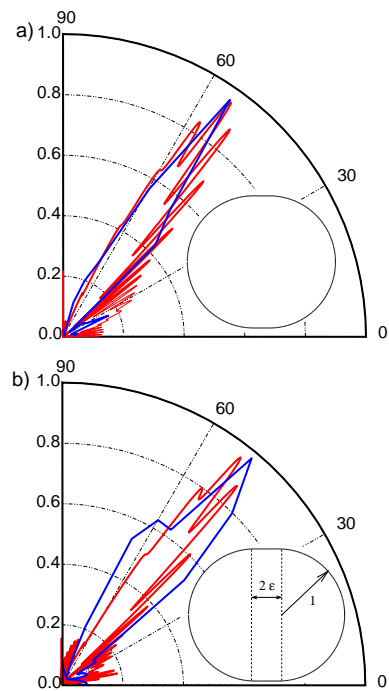}
\caption{Far-field emission patterns for the stadium with $\varepsilon=0.12$ (a) and $\varepsilon=0.18$ (b). The blue curve is the ray simulation and the red a numerical solution of the wave equation; no experimental data was taken for this shape. The ray simulation was performed with random initial conditions exactly as in Fig.~\ref{fig:FFquad}. The numerical solutions were for resonances with $kR=50.5401 -0.0431i$ with $Q=-2\RE[kR]/\IM[kR]=2342.71$ and $kR= 48.7988 -0.1192i $ with $Q=-2\RE[kR]/\IM[kR]=818.83$ for $\varepsilon=0.12$ and $0.18$ respectively. The inset shows the shape of the stadium; it is defined (inset (b)) by two half circles with radius one and a straight line segment of length $2\varepsilon$.}
\label{fig:FFstadium}
\end{figure}

As a further test of the explanatory power of plotting the unstable manifolds, we can use this method to explain the large shift in the far-field directionality in the quadrupole-hexadecapole (QHD) with deformation shown in Fig.~\ref{fig:ExpData}. The QHD shape is an interesting contrast to the quadrupole as the diamond and rectangle periodic orbits interchange their roles. For the QHD at small deformations the diamond orbit is unstable and gives no islands whereas the rectangle is stable and gives rather large islands at $\varepsilon = 0.12$ (see Fig.~\ref{fig:IntensityOktu}a)).  Since there is no island at the point of highest curvature for this deformation the original adiabatic theory would predict emission from the point of highest curvature approximately in the tangent ($90^{\circ}$) direction. We see in Fig.~\ref{fig:FFoktu} that we indeed have such behavior experimentally. The same prediction would come from looking at the unstable manifolds; in Fig.~\ref{fig:IntensityOktu}a) we find the maximum ray escape amplitude comes from near $\phi=0$ and is bounded by the unstable manifold of the diamond orbit.  The relevant unstable manifolds rearrange as the deformation is increased to $\varepsilon =0.20$. The stable rectangle bifurcates at $\varepsilon\sim0.1115$ into two stable, period four, ``parallelogram'' orbits.  As the deformation increases the islands associated with these orbits move closer to $\phi=0$, and although the islands themselves are quite small, they cause the unstable manifolds of the diamond to become steeper (as the fixed point of the diamond orbit sits right between the two period four islands). Due to this the other manifolds also become steeper and at $\varepsilon=0.20$ Fig.~\ref{fig:IntensityOktu}b) shows that the unstable manifold of the rectangular orbit dictates the flow of the escaping rays, which now are emitted into the $\theta=30^\circ$ direction in the far-field for essentially the same reason as in the quadrupole.  Thus the QHD is a shape which behaves like the ellipse at low deformations and as the quadrupole at high deformations; this can be attributed to the evolution in the geometry of its unstable manifolds.
\begin{figure}[!t]
\centering
\includegraphics[width=0.8\linewidth]{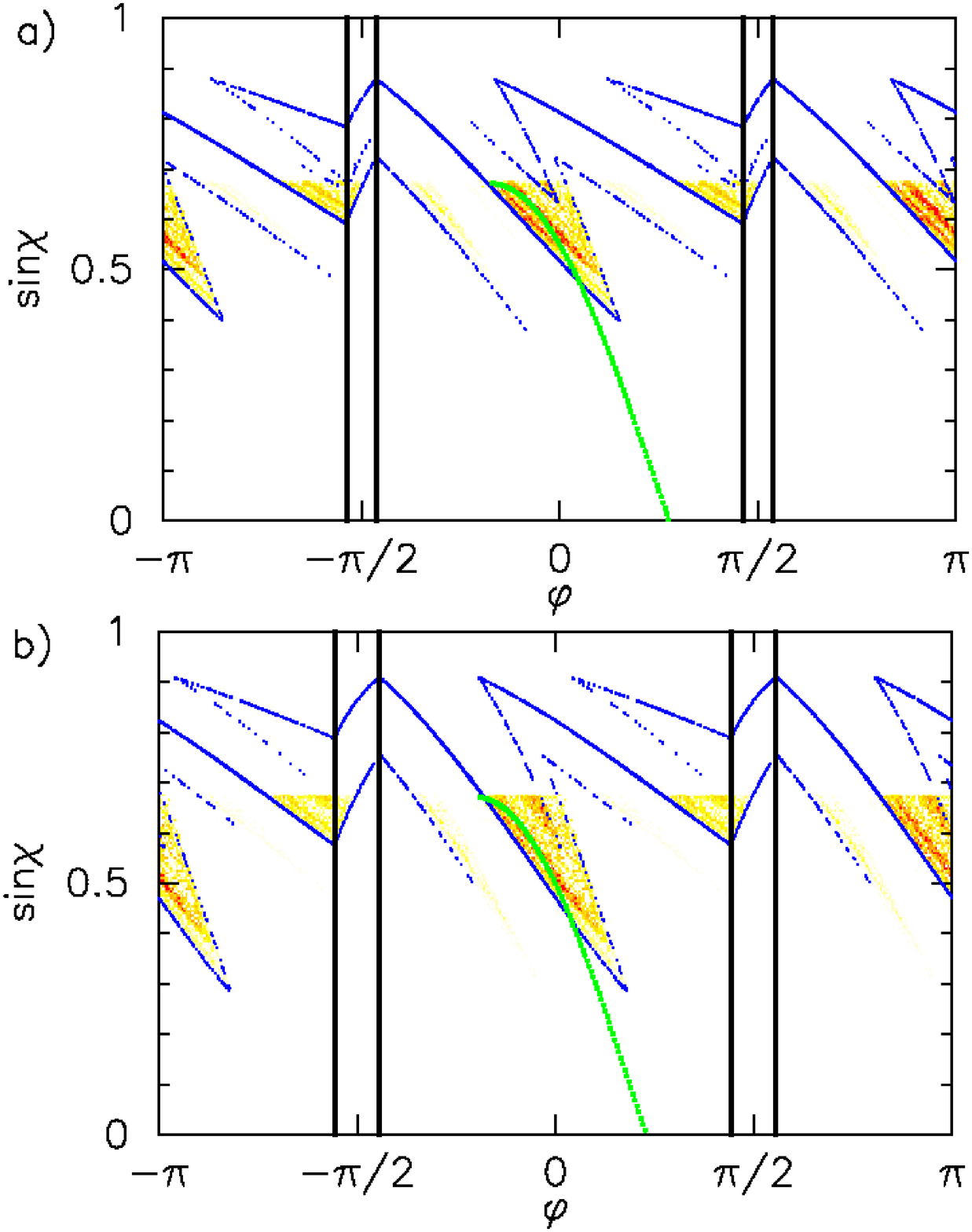}
\caption{Ray emission amplitude (color scale) overlaid on the surface of section for the stadium with $\varepsilon=0.12$ (a) and $\varepsilon = 0.18$ (b). Solid blue curve is the unstable manifold of the periodic rectangle orbit. The green curve is the line of constant $55^\circ$ (a) and $48^\circ$ (b) emission direction into the far-field. The thick black lines mark the end of the circle segments of the boundary and coincide with discontinuities in the manifolds.}
\label{fig:IntensityStadium}
\end{figure}

\section{Directional Emission from Completely Chaotic Resonators}
\label{sect:stadium}

The existence of highly directional emission for the highly deformed quadrupole ($\varepsilon = 0.20$) suggests that the slow diffusion in phase space, characteristic of mixed systems, is not essential to get this effect.  Therefore we decided to study theoretically resonators for which the corresponding billiard is completely chaotic and for which there exist no stable periodic orbits at all.  The Bunimovich Stadium (see inset in Fig.~\ref{fig:FFstadium}), mentioned above, was a natural choice due to its similarity to the quadrupole.  As before we did both ray escape simulations and numerical solutions of the wave equation. In Fig.~\ref{fig:FFstadium} we show our predictions. We find again highly directional emission with a peak direction ($\sim55^\circ$) slightly shifted from the quadrupole; the narrowness of the far-field peak in the stadium is comparable to that of the far-field peak in the quadrupole. We can associate this peak with the slope and position of the manifold of the unstable rectangular orbit in the stadium, Fig.~\ref{fig:IntensityStadium}a). The noticable shift between the $\varepsilon=0.12$ and $\varepsilon=0.18$ deformation (see inset in Fig.~\ref{fig:FFstadium}) originates from the change in the slope of the unstable manifold of the rectangular orbit, Fig.~\ref{fig:IntensityStadium}b). The discontinuities of slope in the unstable manifolds of the periodic orbits in the stadium result from its non-smooth boundary.  These results indicate clearly that a fully chaotic dielectric resonator can nonetheless sustain highly directional lasing modes.  It would be interesting to test this in future experiments.

\section{Summary and conclusions}

We have reported experimental data from polymer micro-pillar lasers with different deformations of the cross-section from circular symmetry. The far-field emission patterns were anisotropic and in most cases highly directional. The anisotropy was stable and reproducible and was dramatically different for different boundary shapes of similar major to minor axis ratio. These differences were reproduced by ray and wave solutions of the ideal passive cavity. The differences in the emission patterns were explained by reference to the different dynamics of rays trapped within each resonator. The possibility of highly directional emission from quadrupole resonators with partially chaotic ray dynamics was predicted earlier, but was found to be much more robust in the experiment than previously suspected. The earlier adiabatic model\cite{NockelS97} could not explain this robustness and a new model based on the geometry of unstable manifolds was proposed and tested. It was able to explain the current data and by extension predicts that fully chaotic resonator shapes such as the stadium would also exhibit highly directional emission.
\section*{Acknowledgment}
We would like to thank Martina Hentschel for her initial contribution to the ray-tracing program. We also would like to thank Grace Chern for numerous fruitful discussions.
NBR and RKC would like to acknowledge the partial support to Yale University from AFOSR (F33615-02-2-6066) and the DARPA SUVOS program under SPAWAR Systems Center (Contract No, N66001-02-8017). ADS would like to acknowledge support from NSF grant DMR-0084501.

\end{document}